\documentclass{article}
\usepackage{spconf,amsmath,graphicx,hyperref}
\usepackage{booktabs}
\usepackage{cite, multirow, makecell, subcaption}

\makeatletter
\newcommand{\unmarkedfootnote}[1]{%
  \begingroup
  \renewcommand\thefootnote{} 
  \footnotetext{#1}%
  \makeatother
  \endgroup
}

\title{CASTELLA: Long Audio Dataset with Captions \\ and Temporal Boundaries}
\name{
    Hokuto Munakata$^{1}$,
    Takehiro Imamura$^{2,*}$,
    Taichi Nishimura$^{1}$,
    and Tatsuya Komatsu$^{1}$
}
\address{LY Corporation$^{1}$, Nagoya University$^{2}$}
\begin{document}
%
\maketitle

\begin{abstract}
We introduce CASTELLA, a human-annotated audio benchmark for the task of audio moment retrieval (AMR). 
Although AMR has various useful potential applications, there is still no established benchmark with real-world data.
The initial study of AMR trained the models solely on synthetic datasets.
Moreover, the evaluation is based on an annotated dataset of fewer than 100 samples.
This resulted in less reliable reported performance.
To ensure performance for applications in real-world environments, we present CASTELLA, a large-scale manually annotated AMR dataset.
CASTELLA consists of 1009, 213, and 640 audio recordings for training, validation, and test splits, respectively, which is 24 times larger than the previous dataset.
We also establish a baseline model for AMR using CASTELLA.
Our experiments demonstrate that a model fine-tuned on CASTELLA after pre-training on the synthetic data outperformed a model trained solely on the synthetic data by 10.4 points in Recall1@0.7. CASTELLA is publicly available in \url{https://h-munakata.github.io/CASTELLA-demo/}.
\end{abstract}


\section{Introduction}
\unmarkedfootnote{* Work done during an internship at LY Corporation.}
The ability to identify audio events and localize them in time is a crucial aspect of audio analysis. Sound event detection (SED) \cite{Ebbers2020, nam22_interspeech, cornell2024dcase,Schmid2024, harju2025audiotext} is often performed using short audio clips, typically around 10 seconds, focusing on coarse-grained event classes.
Recently, audio moment retrieval (AMR) has been proposed, a task that searches for a specific audio segment called ``audio moment'' relevant to a given text query from an input long audio recording ~\cite{munakata2025language}.
AMR enables users to access their desired moments within long audio streams. For example, a listener may retrieve only the engaging parts of a baseball broadcast or quickly locate unusual events in surveillance recordings, without monitoring hours of ordinary audio. By focusing on portions of interest, AMR reduces both time and cognitive effort associated with reviewing lengthy recordings.

\begin{figure}
    \centering
    \includegraphics[width=0.99\linewidth]{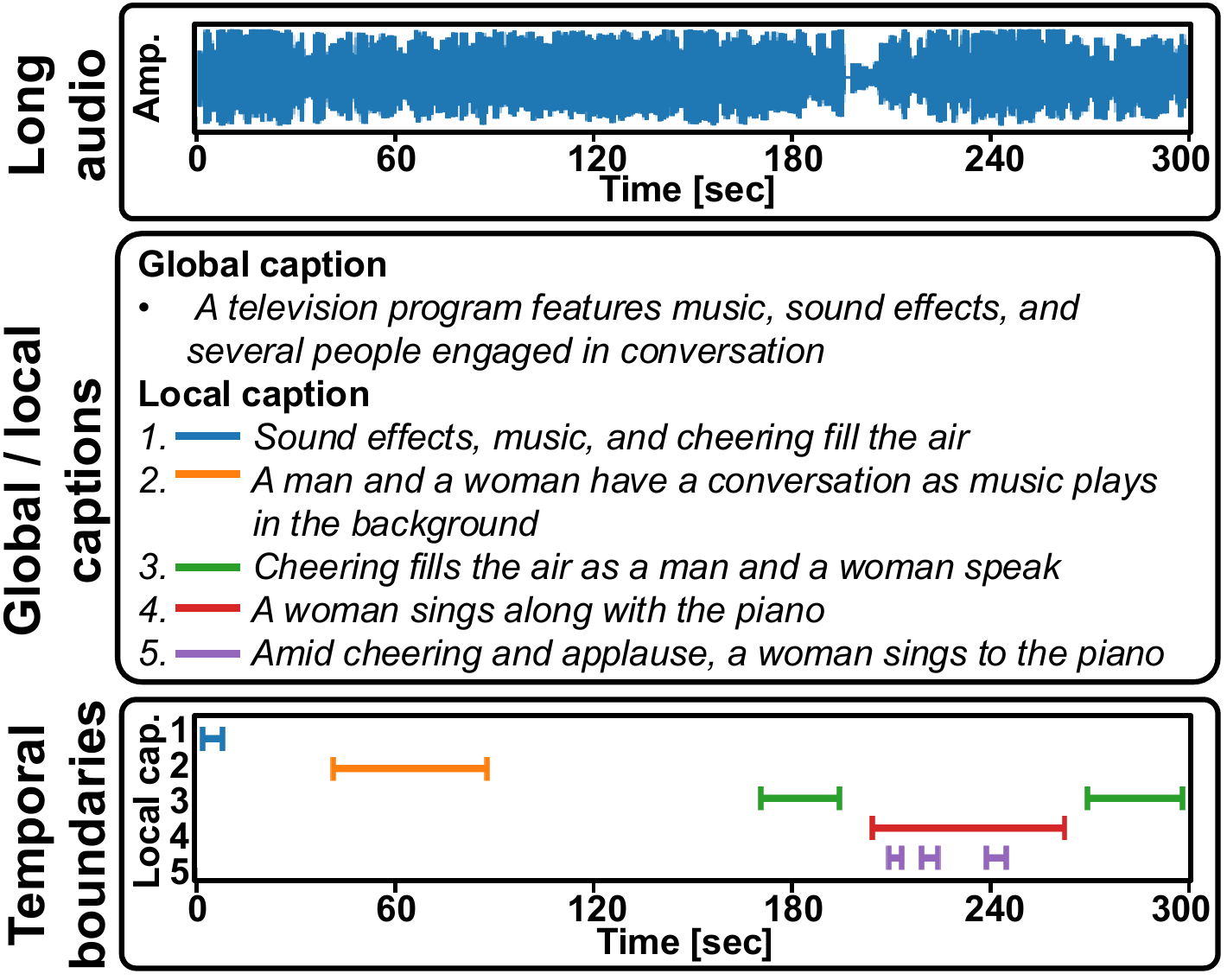}
    \vspace{-3mm}
    \caption{Sample of CASTELLA. Each sample contains long audio recording, global and local captions, and temporal boundaries.}
    \label{fig:sample}
    \vspace{-3mm}
\end{figure}

AMR differs from SED in two key aspects: its handling of long input audio recordings and moments over one minute, and its use of free-format and more complex text queries.
As a specific example of AMR, consider searching for a moment with a highlight piano performance from the input audio in Fig. \ref{fig:sample}.
Given the text query ``A woman sings along with the piano,'' the model predicts the red moment (205s to 264s). In another case, if we are searching for the moments when the performance builds to a climax, an input query could be like ``Amid cheering and applause, a woman sings to the piano'' and the model would predict multiple relevant moments corresponding to the purple ones.

\begin{table*}[t]
    \centering
    \caption{Features of existing audio datasets and CASTELLA.}
    \label{tab: stat1}
    \vspace{-3mm}
    \scalebox{0.95}{
        \begin{tabular}{c|c|c|c|r|c|r|r}
        \toprule
        \multirow{2}{*}{Dataset} & \multirow{2}{*}{\makecell{\makecell{Manual \\ annotation}}} &  \multirow{2}{*}{\makecell{\makecell{Temporal \\ boundary}}} & \multirow{2}{*}{\makecell{Free-format \\ caption}}  & \multirow{2}{*}{\makecell{Total audio len. \\ (hour)}} &  \multirow{2}{*}{\makecell{Audio clip len. \\ (sec)}}  & \multirow{2}{*}{\makecell{\# annotated \\ audio}} & \multirow{2}{*}{\makecell{\# vocab. \\/ label}} \\
        & & & & & & \\
        \midrule
        AudioSet Strong~\cite{hershey2021benefit}     & $\surd$ & $\surd$ & -      & 334  & 10             & 120.5k & 527 \\
        MAESTRO Real~\cite{martin2023strong}            & $\surd$ & $\surd$& -       & 5    & 180-300   & 75   & 8 \\
        AudioCaps~\cite{kim2019audiocaps}           & $\surd$  & -       & $\surd$ & 142  & 10             & 51.3k & 4.8k \\
        TACOS~\cite{primus2025tacos}               & $\surd$ & $\surd$ & $\surd$  & 77   & 10-30     & 12.4k  & 5.2k \\
        Clotho-Moment~\cite{munakata2025language}       & -        & $\surd$ & $\surd$ & 853  & 60             & 44.3k & 4.5k \\
        UnAV-100 subset~\cite{munakata2025language}       & $\surd$ & $\surd$ & $\surd$       & 1  & 45-60             & 77  & 194 \\
        \midrule
        CASTELLA                    & $\surd$ & $\surd$ & $\surd$ & 120  & 60-300  & 1.9k   & 1.8k   \\
        \bottomrule
        \end{tabular}
    }
    \vspace{-5mm}
\end{table*}

While the initial AMR study~\cite{munakata2025language} demonstrated that models can be trained on synthetic datasets, this task lacks a large-scale benchmark consisting of real-world data, which undermines the reliability of the evaluation.
In fact, this study evaluated models on fewer than 100 manually annotated audio recordings.
For the training, this study used synthetic audio constructed by leveraging audio-text pair dataset~\cite{drossos2020clotho} and long background noise~\cite{venkataramanan2023imagenet}.
Consequently, the impact of training on a real-world dataset remains unknown.
To enhance the reliability of the evaluation and demonstrate the effectiveness of training on real-world data, a large-scale human-annotated AMR dataset is necessary.
In addition to large-scale human annotation, building a more desirable AMR benchmark must satisfy the following three requirements: (i) long audio recording that contains at least one audio moment to search for; (ii) free-format captions for these audio moments; and (iii) start and end timestamps of each audio moment.
While several audio datasets for other tasks, such as SED and conventional audio retrieval~\cite{yusong2023laionclap, elizalde2024natural, niizumi24_interspeech, fang25c_interspeech} exist, no existing datasets meet all of the requirements.

In this paper, we propose a new manually annotated audio dataset named CAptionS and TEmporaL boundaries for Long Audio (CASTELLA).
The dataset was built using a crowd-sourcing service, resulting in 1,862 audio recordings with 3,881 captions, which is significantly larger than the previous AMR dataset~\cite{munakata2025language}.
The sample of CASTELLA is shown in Fig~\ref{fig:sample}.
CASTELLA also satisfies the requirements of AMR: (i) It contains long audio recordings of one to five minutes collected from YouTube; (ii) Each audio recording has two types of free-format captions, a global caption and local captions. Global caption summarizes the entire audio recording, and local captions describe salient audio moments occurring locally; and (iii) The local captions have start and end timestamps as moment boundaries.
Based on the constructed dataset, we conducted experiments on the AMR task to establish a baseline model.
Our experiments demonstrate three key insights. First, fine-tuning a model with CASTELLA after pre-training on a synthetic dataset leads to a significant performance improvement.  Second, the performance trend across different architectures is similar to one of video moment retrieval. Finally, accurately retrieving short moments remains a significant challenge.

\begin{table}[t]
    \centering
    \small
    \caption{Statistics of CASTELLA.}
    \vspace{-3mm}
    \label{tab:stat2}
    \begin{tabular}{c|rrr}
    \toprule
    Split & \# audio & \# local captions & \# timestamps \\
    \midrule
    
    Train   & 1009    & 2182    & 6160    \\
    Valid   & 213    & 352    & 973    \\
    Test    & 640    & 1347    & 4175    \\
    \midrule
    Total   & 1862    & 3881    & 11308   \\
    \bottomrule
    \end{tabular}
    \vspace{-3mm}
\end{table}

\section{Audio Moment Retrieval}
Audio moment retrieval (AMR) is the task of searching for an audio moment relevant to input text query from a long audio recording~\cite{munakata2025language}.
Let the input audio be denoted as $\mathbf{x}$, the text query as $q$, and the output audio moments $y$. 
Here, $y$ may consist of multiple moments $y = (y_1, y_2, \ldots, y_N)$ and each moment is represented as a tuple of start and end times, $y_n = (t_{n,\text{start}}, t_{n,\text{end}})$. 
Using these notations, the prediction of an AMR model $f_\text{AMR}$ is expressed as $y \simeq f_\text{AMR}(\mathbf{x}, q)$.
At the same time, the AMR model predicts confidence score $s = (s_1, s_2, \ldots, s_N)$ for each moment to rank the retrieved results.
The training of the AMR model uses pairs of long audio, captions, and temporal boundaries to align the predicted moments with the ground truth, and predict the confidence score correctly.

\section{Features of CASTELLA}
In this section, we first explain existing audio datasets, then compare CASTELLA with them, and provide detailed statistics.

\begin{figure}[t]
    \begin{minipage}[t]{0.47\linewidth}
    \centering
    \includegraphics[width=\linewidth]{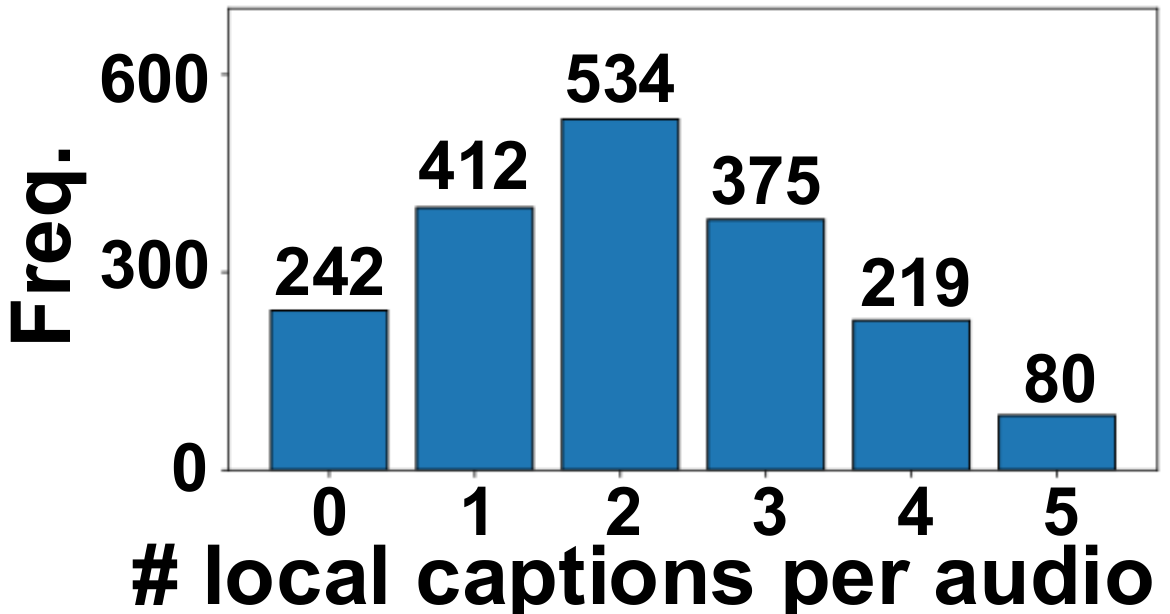}
    \end{minipage}
    \hfill
    \begin{minipage}[t]{0.47\linewidth}
    \centering
    \includegraphics[width=\linewidth]{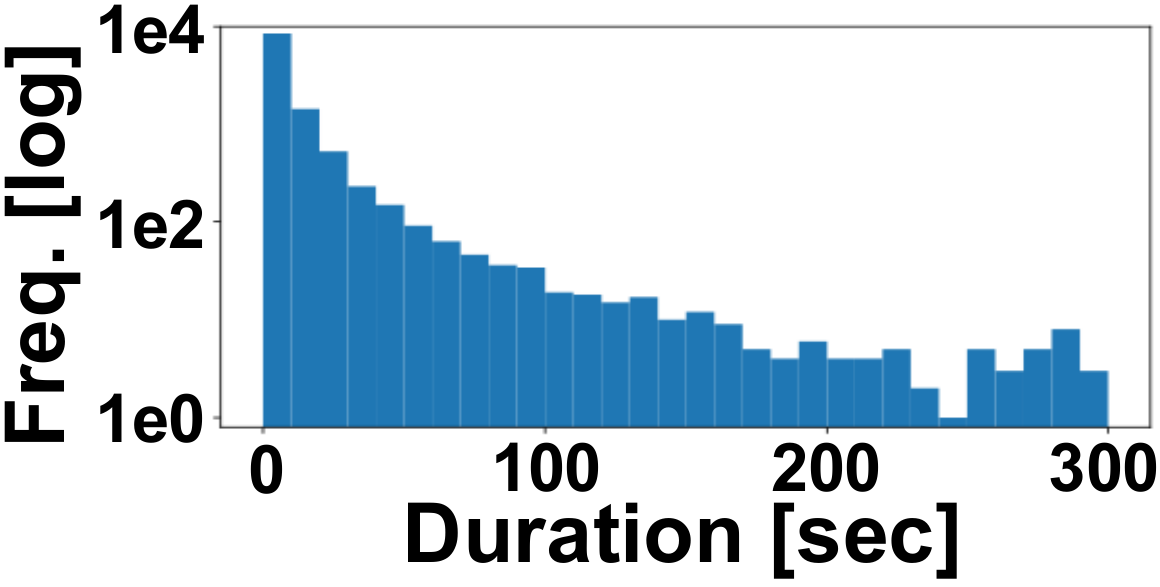}
    \end{minipage}
    \vspace{-3mm}
    \caption{Distributions of the number of local captions per audio (left) and length of timestamps (right).}
    \label{fig: dist}
    \vspace{-3mm}
\end{figure}

\subsection{Existing Audio Dataset}
Table~\ref{tab: stat1} describes the features of existing audio datasets and CASTELLA.
SED datasets have event labels and temporal boundaries~\cite{mesaros2016tut, turpault2019sound, hershey2021benefit, martin2023strong}.
AudioSet strongly labeled subset~\cite{hershey2021benefit} (row 1) provides temporal boundaries as timestamps for a subset of original AudioSet~\cite{gemmeke2017audioset} consisting of 10-second audio clips. This dataset has the largest number of labels and samples.
In contrast, MAESTRO Real~\cite{martin2023strong} (row 2) has audio recordings longer than three minutes, whereas the number of labels is only eight and the amount of data is limited.
Audio-text pair datasets~\cite{drossos2020clotho, mei2024wavcaps, sun2024auto, bai2025audiosetcaps} such as AudioCaps~\cite{kim2019audiocaps} (row 3) have captions freely describing the audio content, but do not have temporal boundaries.
TACOS~\cite{primus2025tacos} (row 4) is audio-text pair dataset also containing temporal boundaries, and the length of the audio clips is ranging from 15 to 30 seconds.
The previous work of AMR proposed Clotho-Moment~\cite{munakata2025language} (row 5), a synthetic AMR dataset generated by overlaying sounds based on the Walking Tours~\cite{venkataramanan2023imagenet} and Clotho ~\cite{drossos2020clotho} dataset.
This work also constructed a manually annotated dataset for the evaluation using subset of UnAV-100~\cite{geng2023dense} (row 6) containing fewer than 100 audio recordings. The length of audio clips in this dataset is less than one minute, which can hardly be considered sufficient.

\subsection{Detailed Statistics of CASTELLA}
CASTELLA is a human-annotated dataset consisting of long audio recordings that range from one to five minutes, free-format captions, and temporal boundaries.
The total length of audio is over 120 hours and the annotated audio recordings are over 1.9k, which is significantly larger than UnAV-100 subset, the previous manually annotated AMR dataset~\cite{munakata2025language}.

The detailed statistics of CASTELLA are summarized in Table~\ref{tab:stat2}.
We collected 1009, 213, and 640 audio recordings for training, validation, and test splits, respectively.
Audio recordings have an average of 2.1 local captions, and local captions have an average of 2.9 timestamps.
Fig.~\ref{fig: dist} (left) shows the distribution of the number of local captions per audio recordings.
Audio recordings with two moments have the highest frequency of occurrence.
Fig.~\ref{fig: dist} (right) shows the distribution of the length of timestamps.
This distribution has a clear long-tail and indicates that short timestamps are dominant.
The vocabulary size is about 1,762 words for both local and global captions, and 1,371 words for local captions alone.
The average length of captions is 7.8 and 13.7 words for local and global captions, respectively.

\section{Building CASTELLA}
In this section, we describe how we built CASTELLA.

\subsection{Collection of Audio Recordings}
We collected long audio recordings from YouTube, specifically from a subset of the AudioCaps split. 
This split also helps avoid contaminating the test split with audio recordings used to train audio-text models.~\cite{yusong2023laionclap, elizalde2024natural}.
Note that AudioCaps selected 75 classes when sampling audio recordings from AudioSet to exclude sounds that are difficult to describe without visual information or expert knowledge; CASTELLA inherits this selection.

Since some audio recordings selected in AudioCaps were too short, we filtered out audio shorter than one minute. 
On the other hand, because extremely long audio is difficult to handle during both training and evaluation, we trimmed the audio recordings to the first five minutes.

\subsection{Annotation Procedure Using Crowd-Sourcing}

We used crowd-sourcing to collect captions and temporal boundaries.
Annotators performed the following three tasks: (1) selecting salient local audio moments; (2) describing global and local captions; (3) translating captions; and (4) assigning temporal boundaries.
Throughout the annotation process, annotators watched videos corresponding to the collected audio. This enables efficient and accurate annotation using visual information.
To improve the quality, a second annotator reviewed the results. If a clear error was found, we asked the original annotator to retry. Additionally, we reviewed all captions and temporal boundaries ourselves to remove errors and visual biases introduced by watching the corresponding videos. We carefully listened to the sound without watching the videos and removed captions that could not be identified from audio information alone or were too quiet to confirm, and we corrected errors in the temporal boundaries.

\noindent\textbf{Selecting salient audio moments}.
Before assigning local captions and temporal boundaries, it is necessary to focus on the prominent local audio moments.
We instructed annotators to watch the entire video and then identify up to five moments that stand out. We also asked them to avoid selecting non-local moments that occur throughout the entire audio recording.
The selected moments can partially overlap, for example, ``A woman sings along with the piano'' and ``Amid cheering and applause, a woman sings to the piano,'' as shown in Fig.~\ref{fig:sample}.

\noindent
\textbf{Collection of global and local captions}.
The captions of the selected audio moments from the previous procedure are described as local captions.
We instructed annotators to describe these moments in as much detail as possible without using visual information. To prevent a loss of caption diversity, we allowed for descriptions that could be inferred from preceding and following sounds, even if the sound source was difficult to determine from the sound alone.
For example, if a horse neighs followed by a scraping sound, it is acceptable to describe it as ``a horse's footsteps.''
After describing the local captions, the annotators then created a summary, referred to as a global caption, considering all the local captions.

\noindent
\textbf{Translating captions.}
While our target language was not only Japanese but also English, it was difficult to find several native English speakers for annotation.
Hence, Japanese captions were initially collected and subsequently translated into English by experienced translators. To expedite the translation process, we provided translators with machine translation results from \texttt{llm-jp-3.1-13b-instruct4}\footnote{https://huggingface.co/llm-jp/llm-jp-3.1-13b-instruct4} and OpenAI's \texttt{GPT-4.1}\footnote{https://openai.com/index/gpt-4-1/}. If the machine translation was appropriate, they were allowed to use it directly.

\noindent
\textbf{Collection of temporal boundaries}.
After collecting captions for the salient audio moments, we also collected their corresponding temporal boundaries. We collected these temporal boundaries with a one-second resolution, as real-world AMR applications do not require a finer resolution.
We instructed annotators to mark timestamps that correspond to or capture more than just the moments described in the local captions. This instruction was based on the idea that for AMR applications, it is desirable to include search results even if they overlap somewhat with non-relevant audio content.
Since we instructed that all corresponding ranges within the audio be annotated, multiple temporal boundaries may be obtained for a single query.

\begin{table*}[t]
    \centering
    \small
    \caption{Retrieval performance on CASTELLA. R1@$\theta$ and mAP@$\theta$ represent recall1 and mean average precision with threshold $\theta$. Higher is better for both metrics.}
    \vspace{-3mm}
    \label{tab:result}
        \begin{tabular}{c|c|cc|rrrrr}
        \toprule
        \multirow{2}{*}{Index}& \multirow{2}{*}{\makecell{DETR-based \\network}} & \multicolumn{2}{c|}{Training dataset} & \multicolumn{5}{c}{AMR Performance} \\
        & & Clotho-Moment & CASTELLA & R1@0.5 & R1@0.7& mAP@0.5 & mAP@0.75 & mAP@avg. \\
        \midrule
        
        1 & \multirow{3}{*}{QD-DETR~\cite{moon2023query}} & $\surd$  & -   & 10.3 & 5.8 & 9.9 & 4.7 & 5.3 \\
        2 & & -  & $\surd$  & 19.8 & 9.7 & 17.6 & 5.9 & 7.7  \\
        3 & &  $\surd$  & $\surd$  & 30.6 & 16.2 & 26.5 & 12.2 & 13.7\\
        \midrule
        4 & Moment-DETR~\cite{lei2021detecting} &  $\surd$  & $\surd$  & 19.3 & 10.8 & 17.2 & 7.0 & 8.2 \\
        5 & UVCOM~\cite{xiao2024bridging}&  $\surd$  & $\surd$  & \textbf{31.7} & \textbf{20.3} & \textbf{28.4} & \textbf{15.2} & \textbf{15.9} \\
        \bottomrule
        \end{tabular}
        \vspace{-5mm}
\end{table*}

\section{Experiment}
In this section, we provide a baseline model for CASTELLA in the AMR tasks.

\subsection{Configuration}
As a baseline, we trained AM-DETR proposed in the previous work~\cite{munakata2025language}.
AM-DETR consists of audio-text model, such as CLAP~\cite{yusong2023laionclap, elizalde2024natural}, with Detection Transformer (DETR)~\cite{carion2020end} inspired by video moment retrieval (VMR) models~\cite{lei2021detecting, moon2023query, jang2023knowing}.
First, the audio-text model extracts audio and text features to capture the multimodal relationship between the input query and audio.
Then, DETR estimates audio moments and their confidence scores by considering inter-frame relationship of audio, with the confidence score used to filter out unnecessary outputs~\cite{carion2020end}.

Following the previous work, we used MS-CLAP~\cite{elizalde2024natural} with sliding window as the feature extractor.
Before the feature extraction, the 242 audio recordings that do not have local captions were removed.
We used a sliding window with both a window length and hop length of one second and downsampled audio recordings to 32 kHz.
We also tested the methods proposed in video moment retrieval. In addition to QD-DETR~\cite{moon2023query}, we used old model named Moment-DETR~\cite{lei2021detecting}~\cite{xiao2024bridging} and UVCOM, an advanced model with improved architecture and learning methods.
The hyperparameters of these DETR-based networks were the same as the original papers.
The optimizer was AdamW~\cite{loshchilov2017decoupled} with a learning rate of \(1 \times 10^{-4}\) and a batch size of 32.
Based on early stopping using validation split, we selected the best model within 100 epochs.
For comparison, we used model weights pre-trained on Clotho-Moment~\cite{munakata2025language}.
All experiments were conducted using a OSS library~\cite{nishimura2024lighthouse} and we will provide the recipe for this experiment upon paper acceptance.

\subsection{Result}
We measured recall1 and mean average precision (mAP) as shown in Table~\ref{tab:result}.
These metrics evaluate whether intersection over union (IoU) between the ground-truth and predicted moments is greater than a given threshold.
Recall1 only evaluate the most confident moment, and mAP considers all predicted moments. The threshold follows the one used in the previous work~\cite{munakata2025language}.

\noindent
\textbf{Training strategy}.
A model trained with CASTELLA using weights pre-trained on Clotho-Moment (Model 3) achieved 16.2 points in recall1@0.7 significantly outperforming models trained using only Clotho-Moment (Model 1) and CASTELLA (Model 2) by 10.4 and 6.5 points, respectively.
This result suggests that pre-training on a synthetic dataset and fine-tuning on a real-world dataset is an effective training strategy.

\noindent
\textbf{Model architecture}.
When comparing the architecture of DETR-based networks, UVCOM (Model 5) performed best, achieving 20.3 points in recall1@0.7, while Moment-DETR (Model 4) was the worst. 
This trend is also observed in the VMR task~\cite{xiao2024bridging}, suggesting that models effective in VMR will also show strong performance in AMR.

\begin{figure}[t]
    \begin{minipage}[t]{0.47\linewidth}
    \centering
    \includegraphics[width=\linewidth]{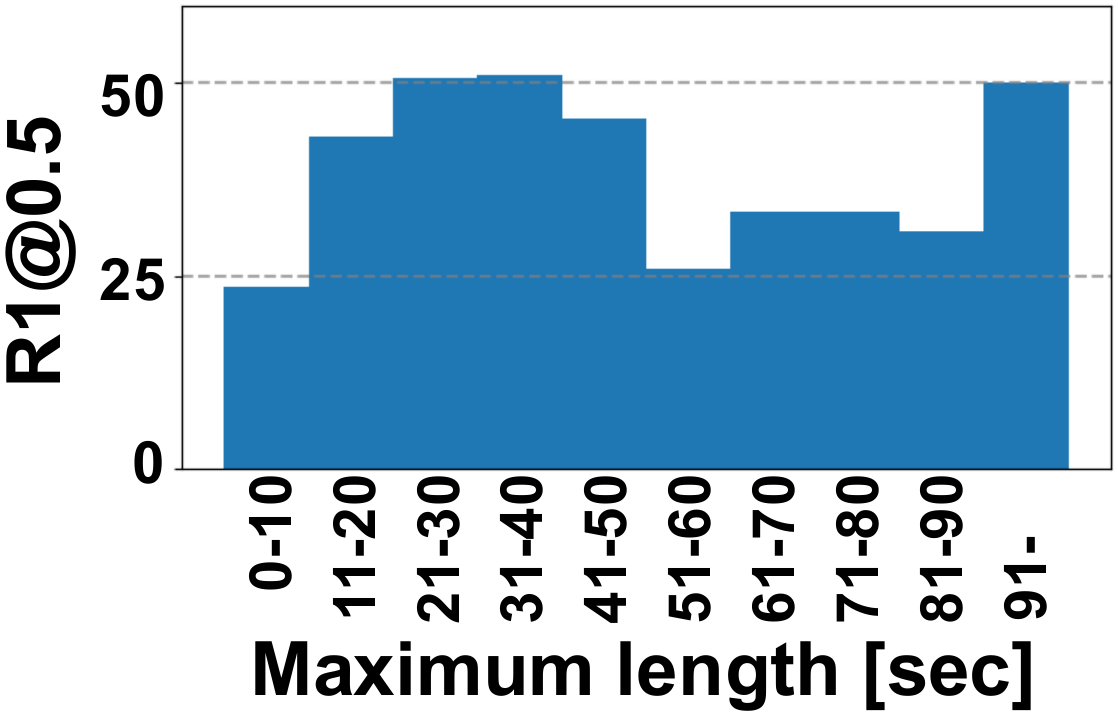}
    \end{minipage}
    \hfill
    \begin{minipage}[t]{0.47\linewidth}
    \centering
    \includegraphics[width=\linewidth]{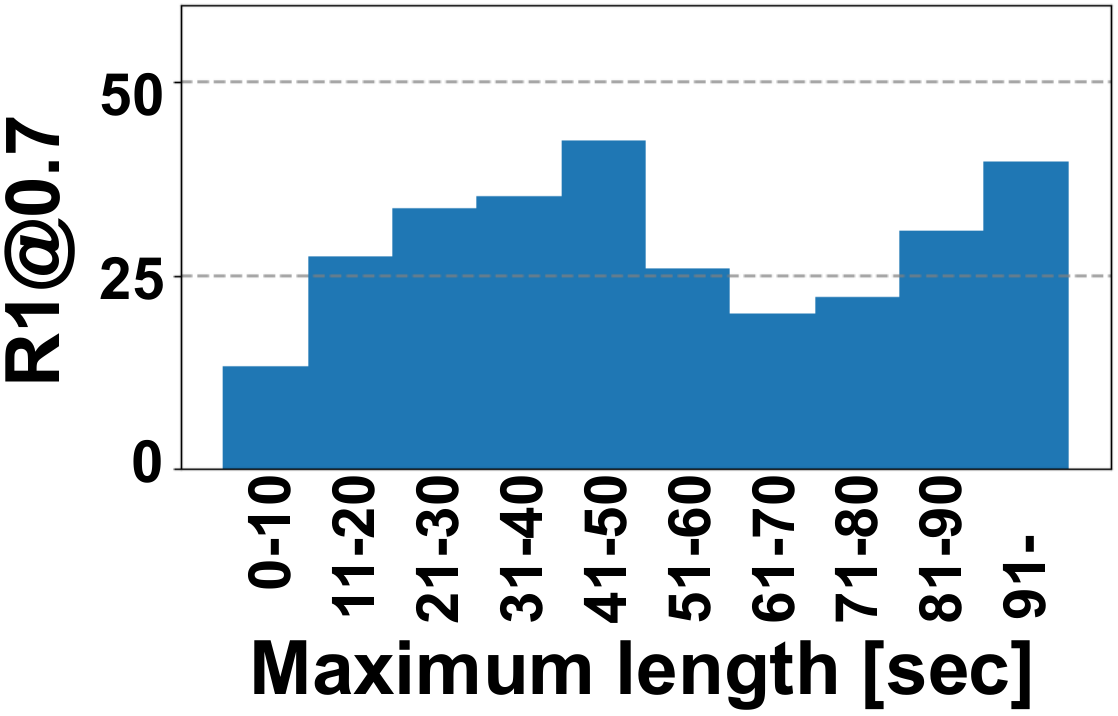}
    \end{minipage}
    \vspace{-3mm}
    \caption{Recall1 by the maximum length of ground-truth audio moments.}
    \label{fig: timestamps}
    \vspace{-3mm}
\end{figure}

\noindent
\textbf{Performance by maximum length of moments}.
Fig.~\ref{fig: timestamps} shows the relationship between recall1 and the maximum length of the ground-truth audio moments.
It can be seen that the performance is low for short moments of less than 10 seconds. This problem has also been reported in VMR~\cite{park2024length}, and it becomes a more serious issue for AMR task because short moments occur frequently in this task as shown in Fig.~\ref{fig: dist}.

\section{Conclusion}
In this paper, we proposed CASTELLA, a manually annotated audio dataset specifically designed for the audio moment retrieval task. 
This dataset contains 1,862 audio recordings, each one to five minutes in length, along with global and local captions, and temporal boundaries collected via crowd-sourcing.
Using CASTELLA, we conducted an AMR experiment with existing models.
Our results show that pre-training on a synthetic dataset and then fine-tuning it on CASTELLA is an effective strategy.

As future work, we will explore new applications for CASTELLA beyond AMR. Specifically, these include developing technologies for the simultaneous localization and captioning of salient events in a long audio recording, as well as methods for retrieving specific audio moments relevant to a query from a collection of multiple long audio recordings.

\section{Acknowledgment}
We appreciate members of LY Communications Corporation reviewing and translating the dataset collected by cloud-sourcing.

\small
\bibliographystyle{IEEEbib}
\bibliography{refs}

\end{document}